\documentclass[twocolumn,pra,showpacs,superscriptaddress,amssymb,amsmath,amsmath]{revtex4-1}
\usepackage{graphicx}
\usepackage{epstopdf}
\usepackage{comment}
\usepackage{hyperref}
\usepackage{bm}

\hypersetup{pdfstartpage=1,pdfmenubar=true,pdfpagemode=None,pdftoolbar=true,
urlcolor=blue,linkcolor=blue,colorlinks = true,citecolor=blue, bookmarksopen=false}

\begin{document}
\title{Ultracold magnetically tunable interactions without radiative charge transfer losses\\ between Ca$^+$, Sr$^+$, Ba$^+$, and Yb$^+$ ions and Cr atoms}

\author{Micha\l~Tomza}
\email{michal.tomza@icfo.es}
\affiliation{ICFO--Institut de Ci\`encies Fot\`oniques, The Barcelona Institute of Science and Technology, 08860 Castelldefels, Spain}

\date{\today}

\begin{abstract}

The Ca$^+$, Sr$^+$, Ba$^+$, and Yb$^+$ ions immersed in an ultracold gas of the Cr atoms are proposed as experimentally feasible heteronuclear systems in which ion-atom interactions at ultralow temperatures can be controlled with magnetically tunable Feshbach resonances without charge transfer and radiative losses. 
\textit{Ab initio} techniques are applied to investigate electronic-ground-state properties of the (CaCr)$^+$, (SrCr)$^+$, (BaCr)$^+$, and (YbCr)$^+$ molecular ions. 
The potential energy curves, permanent electric dipole moments, and static electric dipole polarizabilities are computed.
The spin restricted open-shell coupled cluster method restricted to single, double, and noniterative triple excitations, RCCSD(T), and the multireference configuration interaction method restricted to single and double excitations, MRCISD, are employed. The scalar relativistic effects are included within the small-core energy-consistent pseudopotentials.  
The leading long-range  induction and dispersion interaction coefficients are also reported.
Finally, magnetic Feshbach resonances between the Ca$^+$, Sr$^+$, Ba$^+$, and Yb$^+$ ions interacting with the Cr atoms are analyzed.
The present proposal opens the way towards robust quantum simulations and computations with ultracold ion-atom systems free of radiative charge transfer losses.

\end{abstract}


\maketitle

\section{Introduction}

Hybrid systems of laser-cooled trapped ions and ultracold atoms combined in a single experimental setup are currently emerging as a new platform for fundamental research in quantum physics~\cite{HarterCP14}. Potential applications range from controlled ion-atom collisions and chemical reactions~\cite{ZipkesPRL10,ZipkesNature10,HallPRL11,RatschbacherNatPhys12} to quantum computations~\cite{DoerkPRA10} and simulations of solid-state physics~\cite{GerritsmaPRL12,BissbortPRL13}.
Another possible research direction is the formation of cold molecular ions~\cite{TomzaPRA15a} for precision measurements~\cite{GermannNatPhys14}, ultracold chemistry~\cite{ChangScience13}, or other novel applications~\cite{HarterCP14}.

Losses due to radiative inelastic processes, that is the radiative charge transfer and radiative association~\cite{MakarovPRA03,HallPRL11}, are the basis of controlled ion-atom chemistry~\cite{ChangScience13} but can obstruct sympathetic cooling~\cite{ZipkesNature10}, and interesting applications in quantum computations~\cite{DoerkPRA10} and quantum simulations of many-body physics~\cite{BissbortPRL13}. Radiative losses can be reduced by choosing a system with a favorable electronic structure such as Yb$^+$+Li~\cite{TomzaPRA15a} and avoided if, and only if, an interacting ion-atom system is in its absolute electronic ground state and the charge transfer is impossible. This condition is met when the ionization potential of an atom is larger than the electron affinity of a positive ion.

Up to date, all experimentally investigated cold heteronuclear hybrid ion-atom systems have presented radiative losses~\cite{SmithJMO05,SmithAPB14,HallPRL11,HallMP13a,HallMP13b,HazePRA15,ZipkesNature10,ZipkesPRL10,RatschbacherNatPhys12,
SullivanPRL12,SchmidPRL10,RellergertPRL11,RatschbacherPRL13}. 
These losses are unavoidable for almost all alkaline-earth-metal and ytterbium ions immersed in a cold gas of alkali-metal atoms~\cite{TomzaPRA15a,KrychPRA11,daSilvaNJP2015,AymarJCP11,McLaughlinJPB14,SayfutyarovaPRA13,LambPRA12} (except the Ba$^+$ ion in a gas of the Li atoms).
They can be, however, avoided for some combinations of alkali-metal ions interacting with alkali-metal atoms~\cite{YanPRA14} or alkaline-earth-metal ions interacting with alkaline-earth-metal atoms~\cite{SullivanPRL12} and for almost all combinations of alkali-metal ions interacting with alkaline-earth-metal atoms (except the Li$^+$ ion in a gas of the Ba atoms) [cf.~Table~\ref{tab:losses}].
Homonuclear ion-atom systems do not suffer from radiative losses but show resonant spin-changing charge-exchange inelastic collisions~\cite{GrierPRL09,RaviNatCommun12,GoodmanPRA15}.
Unfortunately, the preparation and manipulation of laser-cooled closed-shell alkali-metal ions and closed-shell alkaline-earth-metal atoms are experimentally more challenging than using open-shell ones. Furthermore, the existence of magnetically tunable Feshbach resonances predicted for mixtures of closed-shell and open-shell species is still waiting for an experimental confirmation~\cite{ZuchowskiPRL10,BruePRL12}. Therefore, the useful control of ultracold ion-atom interactions with magnetically tunable Feshbach resonances should be expected only when both ions and atoms are of the open-shell nature~\cite{JulienneRMP10}.

\begin{table*}[th!]
\caption{The hybrid heteronuclear ion-atom systems without charge transfer radiative losses among the combinations of the present experimentally accessible laser-cooled trapped ions and ultracold atoms.} 
\begin{ruledtabular}
\begin{tabular}{ll}
Nature of ion/atom &  Systems without charge transfer radiative losses   \\
\hline
open-shell/open-shell 
&  Ba$^+$/Li, \textbf{Ca$^+$/Cr}, \textbf{Sr$^+$/Cr}, \textbf{Ba$^+$/Cr}, \textbf{Yb$^+$/Cr}, Sr$^+$/Dy, Ba$^+$/Dy, Sr$^+$/Er, Ba$^+$/Er\\
open-shell/closed-shell 
& Ca$^+$/Mg, Sr$^+$/Mg, Ba$^+$/Mg, Yb$^+$/Mg, Sr$^+$/Ca, Ba$^+$/Ca, Ba$^+$/Sr, Ca$^+$/Yb, Sr$^+$/Yb, Ba$^+$/Yb\\
closed-shell/open-shell 
& Na$^+$/Li, K$^+$/Li, Rb$^+$/Li, Cs$^+$/Li, K$^+$/Na,  Rb$^+$/Na, Cs$^+$/Na, Rb$^+$/K, Cs$^+$/K, Cs$^+$/Rb  \\
& alkali-metal ion/Cr, alkali-metal ion/Dy, alkali-metal ion/Er \\
closed-shell/closed-shell & alkali-metal ion/alkaline-earth-metal atom (except Li$^+$/Ba)
\end{tabular}
\label{tab:losses}
\end{ruledtabular}
\end{table*}

In the present work, we propose the Ca$^+$, Sr$^+$, Ba$^+$, and Yb$^+$ ions immersed in an ultracold gas of the Cr atoms as systems in which ion-atom interactions at ultralow temperatures can be controlled with magnetically tunable Feshbach resonances without charge transfer and radiative losses. 
Among the present experimentally accessible laser-cooled trapped ions and ultracold atoms, the other possible combinations are the Ba$^+$ ions and the Li atoms as well as the Sr$^+$ or Ba$^+$ ions and the Dy or Er atoms.
The interaction of the ${}^2S$ state alkaline-earth-metal or ytterbium ion with the ${}^7S$ state chromium atom  gives rise to the two molecular electronic states of the $X^6\Sigma^+$ and $a^8\Sigma^+$ symmetries~\cite{Herzberg}. 
Here we report the \textit{ab initio} properties of these electronic states of the (CaCr)$^+$, (SrCr)$^+$, (BaCr)$^+$, and (YbCr)$^+$ molecular ions and prospects for magnetically tunable Feshbach resonances opening the way towards more elaborate studies of the formation and application of the proposed hybrid ion-atom systems.

The plan of our paper is as follows. Section~\ref{sec:theory} describes the theoretical methods used in the electronic structure and scattering calculations. Section~\ref{sec:results} discusses the potential energy curves and electric properties of the studied molecular ions and prospect for magnetically tunable Feshbach resonances. Section~\ref{sec:summary} summarizes our paper.

\section{Computational details}
\label{sec:theory}

\subsection{Electronic structure calculations}

We adopt the computational scheme successfully applied to the ground states of the (LiYb)$^+$ molecular ion~\cite{TomzaPRA15a}, the chromium--alkaline-earth-metal-atom molecules~\cite{TomzaPRA13a}, and the europium--alkali-metal-atom molecules~\cite{TomzaPRA14}.
The potential-energy curves for the $a^8\Sigma^+$ electronic state in the Born-Oppenheimer approximation are obtained with the spin-restricted open-shell coupled cluster method restricted to single, double, and noniterative triple excitations, starting from the restricted open-shell Hartree-Fock (ROHF) orbitals, RCCSD(T)~\cite{KnowlesJCP99}.
The energy splittings between the potential energy curves of the $X^6\Sigma^+$ and $a^8\Sigma^+$  electronic states are calculated with the multireference configuration interaction method restricted to single and double excitations (MRCISD)~\cite{WernerJCP88}.
The potential energy curves of the $X^6\Sigma^+$ electronic ground state are obtained by subtracting the energy splittings from the potential energy curves of the $a^8\Sigma^+$ electronic state calculated with the RCCSD(T) method.
The interaction energies are obtained with the supermolecule method with
the basis set superposition error corrected by using the counterpoise correction~\cite{BoysMP70}
\begin{equation}
V_{(X\textrm{Cr})^+}=E_{(X\textrm{Cr})^+}-E_{X^+}-E_\textrm{Cr}\,,
\end{equation}
where $E_{(X\textrm{Cr})^+}$ denotes the total energy of the molecular ion, and $E_{X^+}$ and $E_\textrm{Cr}$ are the total energies of the ion and atom computed in the diatom basis set. 

The scalar relativistic effects are included by employing the small-core relativistic energy-consistent pseudopotentials (ECP) to replace the inner-shells electrons~\cite{DolgCR12}. The use of the pseudopotentials  allows one to use larger basis sets to describe the valence electrons and models the inner-shells electrons density as accurately as the high quality atomic calculation used to fit the pseudopotentials.
The pseudopotentials from the Stuttgart library are employed in all calculations. The Cr atom is described by the ECP10MDF pseudopotential~\cite{DolgJCP87} and the $[14s13p10d5f4g3h]$ basis set with coefficients taken from the aug-cc-pVQZ-DK basis~\cite{BalabanovJCP05}.
The Ca, Sr, Ba, and Yb atoms are  described with the ECP10MDF, ECP28MDF, ECP46MDF, and ECP28MDF pseudopotentials~\cite{LimJCP06,DolgTCA98} and the $[12s12p7d4f2g]$, $[14s11p6d5f4g]$, $[13s12p6d5f4g]$, and $[15s14p12d11f8g]$ basis sets, respectively, obtained by augmenting the basis sets suggested in Refs.~\cite{LimJCP06,DolgTCA98}.
The basis sets are additionally augmented in all calculations by the set of $[3s3p2d1f1g]$ bond functions~\cite{midbond}.

The interaction potential between the $X^+$($^2S$) ion and the Cr($^7S$) atom, both in the electronic ground state, at large internuclear distances $R$ reads~\cite{KrychPRA11}
\begin{equation}\label{eq:E6}
V_{(X\textrm{Cr})^+}(R)=-\frac{C^\textrm{ind}_4}{R^4}-\frac{C^\textrm{ind}_6}{R^6}-\frac{C^\textrm{disp}_6}{R^6}+\dots\,,
\end{equation}  
where the leading long-range induction coefficients are $C^{\mathrm{ind}}_4=\frac{1}{2}\alpha_\textrm{Cr}$ and $C^{\mathrm{ind}}_6=\frac{1}{2}\beta_\textrm{Cr}$, and  the dispersion one is $C^{\mathrm{disp}}_6=\frac{3}{\pi}\int^\infty_0\alpha_{X^+}(i\omega)\alpha_\textrm{Cr}(i\omega)d\omega$. $\alpha_\textrm{Cr}$ is the static electric dipole polarizability of the Cr atom, $\beta_\textrm{Cr}$ is the static electric quadrupole polarizability of the Cr atom, and $\alpha_{X^+/\textrm{Cr}}(i\omega)$ is the dynamic polarizbility of the $X^+$ ion and the Cr atom at imaginary frequency.  
The dynamic polarizabilities at imaginary frequency of the atom and ions are obtained from the sum over state expression using the transition moments from Refs.~\cite{TomzaPRA13a,SafronovaPRA09,KaurPRA15} and energy levels from the NIST Atomic Spectra Database~\cite{nist}.  
The static quadrupole polarizability of the Cr atom is calculated with the RCCSD(T) and finite field
methods.

The permanent electric dipole moments and static electric
dipole polarizabilities are calculated with the finite field
method using the RCCSD(T) and MRCISD methods for the $a^8\Sigma^+$ and $X^6\Sigma^+$ electronic states, respectively.
The $z$ axis is chosen along the internuclear axis,
oriented from the alkaline-earth-metal ion to the chromium atom, and the origin is set in the center of mass.

All electronic structure calculations are performed with the \textsc{Molpro} package of \textit{ab initio} programs~\cite{Molpro}.

\subsection{Scattering calculations}

The Hamiltonian describing the relative nuclear motion of the $X^+$(${}^2S$) ion and the Cr($^7S$) atom reads
\begin{widetext}
  \begin{equation}\label{eq:Ham}
    \hat{H}=-\frac{\hbar^2}{2\mu}\frac{1}{R}\frac{d^2}{dR^2}R+
    \frac{\hat{l}^2}{2\mu R^2}+
    \sum_{S,M_S}V_S(R)|S,M_S\rangle\langle S,M_S|+V^{ss}(R)+
    \hat{H}_{X^+}+\hat{H}_{\mathrm{Cr}}\,,
  \end{equation}
\end{widetext}
where $\mu$ is the reduced mass, $R$ is the internuclear distance, $\hat{l}$ is the rotational
angular momentum operator, and $V_S(R)$ is the potential energy curve for the
state with total electronic spin $S$. $V^{ss}(R)$ is the spin-dipole--spin-dipole interaction
responsible for the  dipolar relaxation~\cite{StoofPRB88}. 
The atomic Hamiltonian, $\hat{H}_j$ ($j=X^+$, Cr),
including Zeeman and hyperfine interactions, is given by 
\begin{equation}\label{eq:Ham_at}
\hat{H}_j=\zeta_{j}\hat{i}_{j}\cdot\hat{s}_{j}
  +\left(g_e\mu_{{B}}\hat{s}_{j,z}+g_{j}\mu_{{N}}\hat{i}_{j,z}\right)B_z
\end{equation}
where $\hat{s}_{j}$ and $\hat{i}_{j}$ are the electron and
nuclear spin operators, $\zeta_{j}$ is the hyperfine coupling
constant, $g_{e/j}$ are the electron and 
nuclear $g$ factors, and $\mu_{B/N}$ are the Bohr and nuclear
magnetons.  
The experimental values of the nuclear $g$ factors from the NIST Atomic Spectra Database~\cite{nist} and the following hyperfine coupling constant are assumed: $-806.4\,$MHz for $^{43}$Ca$^+$~\cite{ArbesZPD94}, $-1000.5\,$MHz for $^{87}$Sr$^+$~\cite{SunaoshiHI93}, $4018.8\,$MHz for $^{137}$Ba$^+$~\cite{TrapprHI00}, $-3497.2\,$MHz for $^{173}$Yb$^+$~\cite{OlmschenkPRA07}, and $-83.60\,$MHz for $^{53}$Cr~\cite{ReinhardtZPD95}.
Negative values mean that the total angular momentum is larger for the ground hyperfine state as compared to the excited one.
The impact of the Lorentz force and Landau quantization effects~\cite{SimoniJPB11} is neglected in the present study.

The total scattering wave function is constructed in a fully uncoupled
basis set assuming the projection of the total angular momentum to be conserved. 
The coupled channels equations are solved using the renormalized Numerov
propagator~\cite{JohnsonJCP78} with step-size doubling. 
The wave function ratio is propagated to  large interatomic separations, transformed to the diagonal basis, and
the $K$ and $S$ matrices are extracted by imposing the long-range scattering boundary
conditions in terms of the Bessel functions.
The scattering length is obtained from the $S$ matrix for the entrance channel 
$a_0=(1-S_{11})(1+S_{11})/(ik)$, 
where $k=\sqrt{2\mu E/\hbar^2}$ with the collision energy $E$~\cite{IdziaszekPRA09,IdziaszekNJP11,CotePRA00}.

\section{Numerical results and discussion}
\label{sec:results}

\subsection{Potential energy curves}

\begin{figure}[t!]
\begin{center}
\includegraphics[width=\columnwidth]{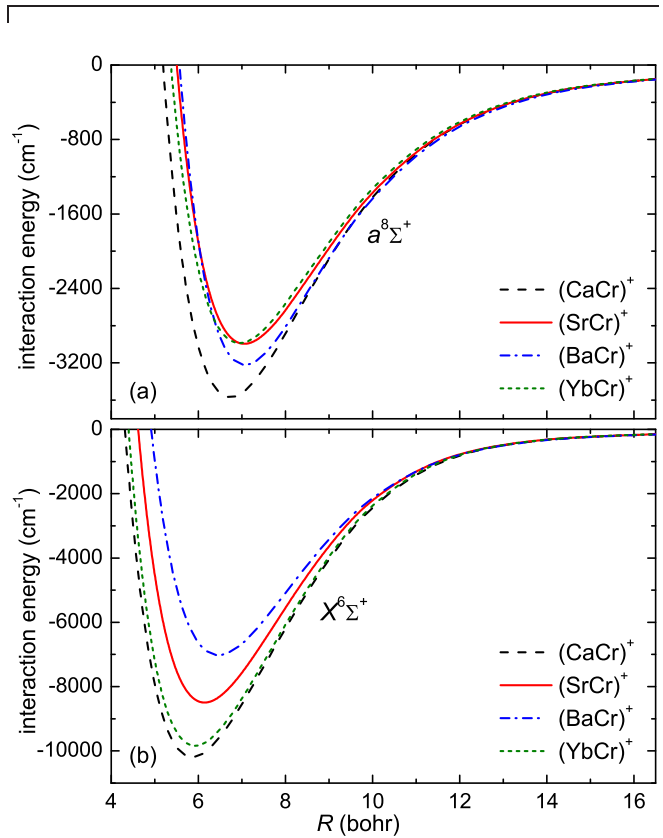}
\end{center}
\caption{(Color online) Potential energy curves of the $a^8\Sigma^+$~(a) and $X^6\Sigma^+$~(b) electronic states of the (CaCr)$^+$, (SrCr)$^+$, (BaCr)$^+$, and (YbCr)$^+$ molecular ions.}
\label{fig:curves}
\end{figure}

\begin{figure}[t!]
\begin{center}
\includegraphics[width=\columnwidth]{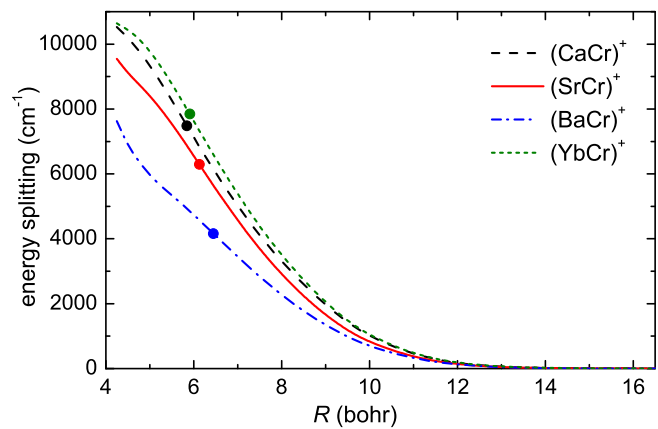}
\end{center}
\caption{(Color online) Energy splittings between the $X^6\Sigma^+$ and $a^8\Sigma^+$ electronic states of the (CaCr)$^+$, (SrCr)$^+$, (BaCr)$^+$, and (YbCr)$^+$ molecular ions. The points indicate the values for the equilibrium distance of the $X^6\Sigma^+$state.}
\label{fig:split}
\end{figure}

The interaction of the ground-state alkaline-earth-metal or ytterbium ion with the ground-state chromium atom, which both are open-shell, results in the two electronic states of the  $a^8\Sigma^+$ and $X^6\Sigma^+$ symmetries. 
The computed potential energy curves for these electronic states of the (CaCr)$^+$, (SrCr)$^+$, (BaCr)$^+$, and (YbCr)$^+$ molecular ions are presented in Fig.~\ref{fig:curves}(a) and Fig.~\ref{fig:curves}(b), respectively, and show a smooth behavior with well-defined minima.
Their equilibrium distances $R_e$ and well depths $D_e$ are reported  in Table~\ref{tab:spec}.
The energy splittings between the $a^8\Sigma^+$ and $X^6\Sigma^+$ states are shown in Fig.~\ref{fig:split}.
The leading long-range induction and dispersion interaction coefficients are reported in Table~\ref{tab:Cn}.

\begin{table*}[tb]
\caption{Spectroscopic characteristics: Equilibrium bond length $R_e$, well depth $D_e$, harmonic frequency $\omega_0$, and number of bound vibrational states $N_v$ of the $X^6\Sigma^+$ and $a^8\Sigma^+$ electronic states and rotational constant $B_0$, electric dipole moment $d_0$, average polarizability $\bar{\alpha}_0$, and polarizability anisotropy $\Delta\alpha_0$, for the rovibrational ground level 
of the $X^6\Sigma^+$ and $a^8\Sigma^+$  electronic states of the (CaCr)$^+$, (SrCr)$^+$, (BaCr)$^+$, and (YbCr)$^+$ molecular ions. 
Masses of the most abundant isotopes are assumed. }
\begin{ruledtabular}
\begin{tabular}{lrrrrrrrr}
System & $R_e\,$(bohr) & $D_e\,$(cm$^{-1}$) &  $\omega_0\,$(cm$^{-1}$) & $N_v$ & $B_0\,$(cm$^{-1}$)&  $|d_0|\,$(D) & $\bar\alpha_0\,$(bohr$^3$) & $\Delta\alpha_0\,$(bohr$^3$)   \\
\hline 
\multicolumn{9}{c}{$\mathbf{a^8\Sigma^+}$} \\
(CaCr)$^+$ & 6.72 & 3582  &  87.7 & 110 & 0.059 & 3.58 & 158 & 88 \\
(SrCr)$^+$ & 7.04 & 3007  &  66.5 & 125 & 0.041 & 1.17 & 176 & 80 \\
(BaCr)$^+$ & 7.06 & 3235  &  67.2 & 136 & 0.032 & 0.48 & 187 & 20 \\
(YbCr)$^+$ & 6.93 & 2998  &  58.9 & 139 & 0.031 & 1.18 & 142 & 81 \\
\multicolumn{9}{c}{$\mathbf{X^6\Sigma^+}$}\\
(CaCr)$^+$ & 5.84 & 10233 & 147.5 & 154 & 0.078 & 3.17 & 126 &  80 \\
(SrCr)$^+$ & 6.13 & 8534  & 115.5 & 171 & 0.049 & 1.22 & 145 & 104 \\
(BaCr)$^+$ & 6.45 & 7051  &  98.3 & 172 & 0.038 & 0.45 & 185 & 157 \\
(YbCr)$^+$ & 5.91 & 9879  & 110.7 & 202 & 0.043 & 2.19 & 104 & 60 \\
\end{tabular}
\label{tab:spec}
\end{ruledtabular}
\end{table*}

\begin{table}[tb]
\caption{Induction and dispersion interaction coefficients describing the
  long-range part of the interaction potential between the alkaline-earth-metal and Yb ions,
  and the Cr atom, all in the electronic ground state.\label{tab:Cn}} 
\begin{ruledtabular}
\begin{tabular}{lrrr}
System &  $C^{\mathrm{ind}}_4\,$(a.u.) & $C^{\mathrm{ind}}_6\,$(a.u.) &   $C^{\mathrm{disp}}_6\,$(a.u.) \\
\hline
Ca$^+$+Cr & 40.8 & 290 & 557  \\
Sr$^+$+Cr & 40.8 & 290 & 654 \\
Ba$^+$+Cr & 40.8 & 290 & 807 \\
Yb$^+$+Cr & 40.8 & 290 & 484 \\
\end{tabular}
\label{tab:Cn}
\end{ruledtabular}
\end{table}

The ion-atom interaction is dominated by the induction contribution, that results in relatively large binding energy of the $X^6\Sigma^+$ electronic ground state. The well depths of this state are between 7051$\,$cm$^{-1}$ for the (BaCr)$^+$ molecular ion and 10233$\,$cm$^{-1}$ for the (CaCr)$^+$ molecular ion. The corresponding equilibrium distances are between 5.84$\,$bohr and 6.45$\,$bohr for the (CaCr)$^+$ and (BaCr)$^+$ molecular ions, respectively.  
The binding energy of the $a^8\Sigma^+$ high-spin electronic state is around three times smaller and its equilibrium distance is around 1$\,$bohr longer as compared to  the $X^6\Sigma^+$ electronic ground state. 
 The well depths of this state are between 2998$\,$cm$^{-1}$ for the (YbCr)$^+$ molecular ion and 3582$\,$cm$^{-1}$ for the (CaCr)$^+$ molecular ion.
 The corresponding equilibrium distances are between 6.72$\,$bohr and 7.06$\,$bohr for the (CaCr)$^+$ and (BaCr)$^+$ molecular ions, respectively. The characteristics of the presented potential  energy curves are similar with a small variation of the binding energy and equilibrium distance   among investigated ion-atom systems. The long-range parts of the interaction potentials are also almost identical [cf.~Fig.~\ref{fig:curves} and Table~\ref{tab:Cn}]. This behavior results from the fact that the electronic properties of the investigated ion-atom systems are determined  by the induction interaction between 
different ions (in the same electronic state) but the same chromium atom.

The $X^6\Sigma^+$ electronic ground state of the investigated molecular ions is bound around three times stronger as compared to the isoelectronic chromium--alkali-metal-atom molecules~\cite{PavlovicPRA10,JeungJPB10}. The stabilization due to the induction interaction is even more pronounced in the case of the $a^8\Sigma^+$ high-spin electronic state which is bound around twenty times stronger in molecular ions as compared to their isoelectronic neutral counterparts. In fact, the potential energy curves for the $a^8\Sigma^+$ high-spin electronic state of the investigated molecular ions are as deep as for the $X^8\Sigma^+$ electronic ground state of the chromium--alkali-metal-atom molecules and the $X^7\Sigma^+$ electronic ground state of the chromium--alkaline-earth-metal-atom molecules.
Interestingly, the energy splittings between the $X^6\Sigma^+$ and $a^8\Sigma^+$ electronic states (dominated by the exchange energy and responsible for standard magnetic Feshbach resonances) of the investigated molecular ions are of the same order of magnitude as in the isoelectronic chromium--alkali-metal-atom molecules.

\begin{figure}[th!]
\begin{center}
\includegraphics[width=\columnwidth]{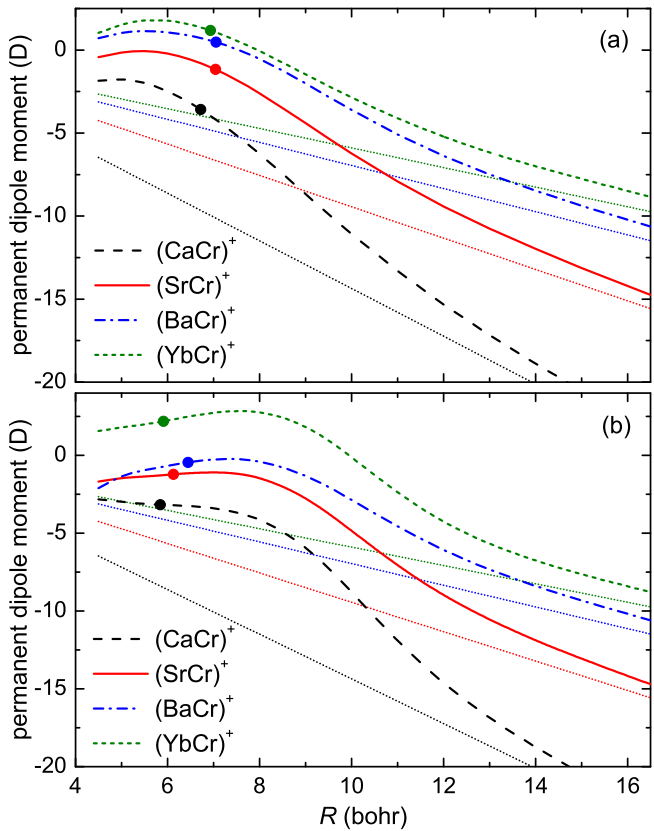}
\end{center}
\caption{(Color online) Permanent electric dipole moment of the (CaCr)$^+$, (SrCr)$^+$, (BaCr)$^+$, and (YbCr)$^+$ molecular ions in the $a^8\Sigma^+$~(a) and $X^6\Sigma^+$~(b) electronic states as a function of the internuclear distance. 
The $z$ axis is oriented from $X^+$ to Cr and the origin is in the center
of mass. The points indicate the values for the ground rovibrational
level, and the dotted lines represent the permanent dipole moment that
the molecular ion would have if the charge was completely localized
on the ion.}
\label{fig:dip}
\end{figure}

The accuracy of \textit{ab initio} calculations depends on the proper treatment of relativistic effects, the
reproduction of the correlation energy, and the convergence in the size of the basis function set.
The employed CCSD(T) method is the golden standard of quantum chemistry and a good compromise between the computational cost and  the accuracy~\cite{MusialRMP07}.
To test the ability of the used \textit{ab initio} approach with the employed basis sets and energy-consistent pseudopotentials to reproduce the experimental data, we check the accuracy of the atomic results.
The predicted ionization potentials are 49243$\,$cm$^{-1}$ for Ca, 45814$\,$cm$^{-1}$ for Sr, 41790$\,$cm$^{-1}$ for Ba, and 50073$\,$cm$^{-1}$ for Yb, in good agreement with the experimental values of 49305$\,$cm$^{-1}$, 45932$\,$cm$^{-1}$, 42035$\,$cm$^{-1}$, and 50443~\cite{nist}, respectively. 
The predicted static electric dipole polarizabilities of the ground state are 76.5$\,a_0^3$ for Ca$^+$, 92.0$\,a_0^3$ for Sr$^+$, 121.7$\,a_0^3$ for Ba$^+$, and 61.6$\,a_0^3$ for Yb$^+$, in good agreement with the experimental values of 75.3$\pm$0.4$\,a_0^3$~\cite{ChangJPB83}, 91.3$\pm0.9\,a_0^3$~\cite{JiangJPB09}, $123.88\pm0.05\,a_0^3$~\cite{SnowPRA07}, and 62.04$\,a_0^3$~\cite{SafronovaPRA09}, respectively, and with the most recent theoretical results~\cite{KaurPRA15}.
The potential well depths of the alkaline-earth-metal and alkali-metal dimers, using the same methodology as in the present study, are reproduced with an error of a few percent as compared to the experimental results~\cite{SkomorowskiJCP12,TomzaMP13}. 
Based on the above and additional convergence analysis,  
we estimate the total uncertainty of the calculated potential
energy curves to be of the order of 100-200$\,$cm$^{-1}$ that corresponds to 3-5\% of the interaction energy. The uncertainty of other electronic properties, including the long-range interaction coefficients, is of the same order of magnitude.

\subsection{Permanent electric dipole moments and static electric dipole polarizabilities}

\begin{figure}[t!]
\begin{center}
\includegraphics[width=\columnwidth]{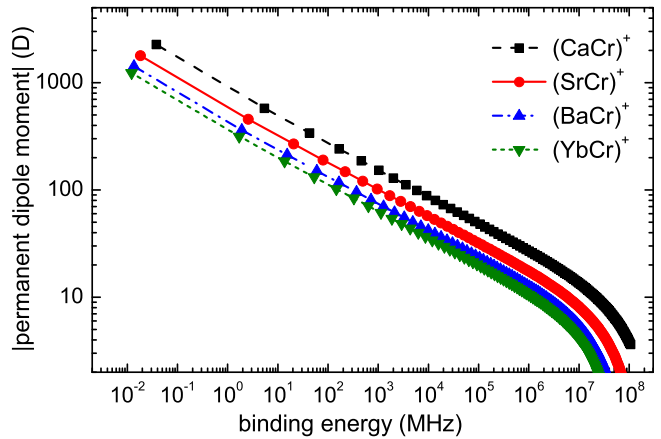}
\end{center}
\caption{(Color online) Absolute values of the permanent electric dipole moment of the (CaCr)$^+$, (SrCr)$^+$, (BaCr)$^+$, and (YbCr)$^+$ molecular ions in different vibrational levels of the $a^8\Sigma^+$ electronic state as a function of the binding energy. The origin is in the center of mass.}
\label{fig:vib_dip}
\end{figure}

A static electric field, that couples with a permanent electric dipole moment and orients molecules, can be used to control and manipulate the dynamics of neutral polar molecules at ultralow temperatures~\cite{QuemenerCR12}. 
Unfortunately, the spatial motion of charged ions is highly sensitive to an external dc electric field and the control of their dynamics with this kind of field is limited. Instead, a nonresonant laser field, that couples with a electric dipole polarizbility and aligns molecules, can be used to tune ion-atom collision by means of the ac Stark shift.
Such control of the photodissociation~\cite{SussmanScience06}, photoassociation~\cite{GonzalezPRA12}, magnetoassociation~\cite{TomzaPRL14}, and rovibrational structure dynamics~\cite{TomzaMP13} has already been investigated in the case of neutral diatomics.

The permanent electric dipole moments of the (CaCr)$^+$, (SrCr)$^+$, (BaCr)$^+$, and (YbCr)$^+$ molecular ions in the  $a^8\Sigma^+$ and  $X^6\Sigma^+$ electronic states as functions of the internuclear distance $R$  are presented in Fig.~\ref{fig:dip}(a) and Fig.~\ref{fig:dip}(b), respectively. The values for the ground rovibrational level are reported in Table~\ref{tab:spec} and the values for all vibrational levels of the $a^8\Sigma^+$ electronic state are shown in Fig.~\ref{fig:vib_dip}. 

\begin{figure}[t!]
\begin{center}
\includegraphics[width=\columnwidth]{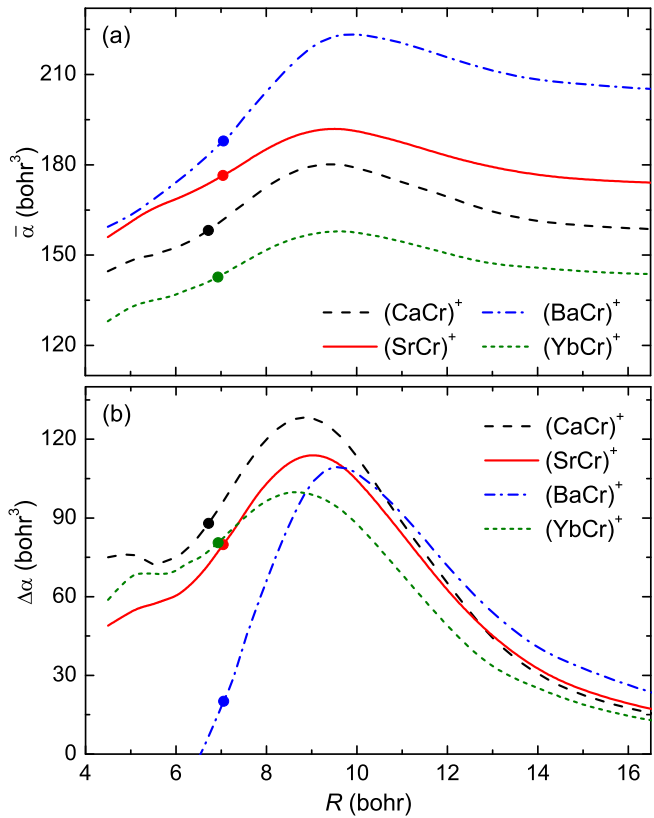}
\end{center}
\caption{(Color online) Average static electric dipole polarizability (a) and polarizability anisotropy (b) of the (CaCr)$^+$, (SrCr)$^+$, (BaCr)$^+$, and (YbCr)$^+$ molecular ions in the $a^8\Sigma^+$ electronic state as a function of the internuclear distance. The points indicate the values for the ground rovibrational level.}
\label{fig:pol_a}
\end{figure}

\begin{figure}[t!]
\begin{center}
\includegraphics[width=\columnwidth]{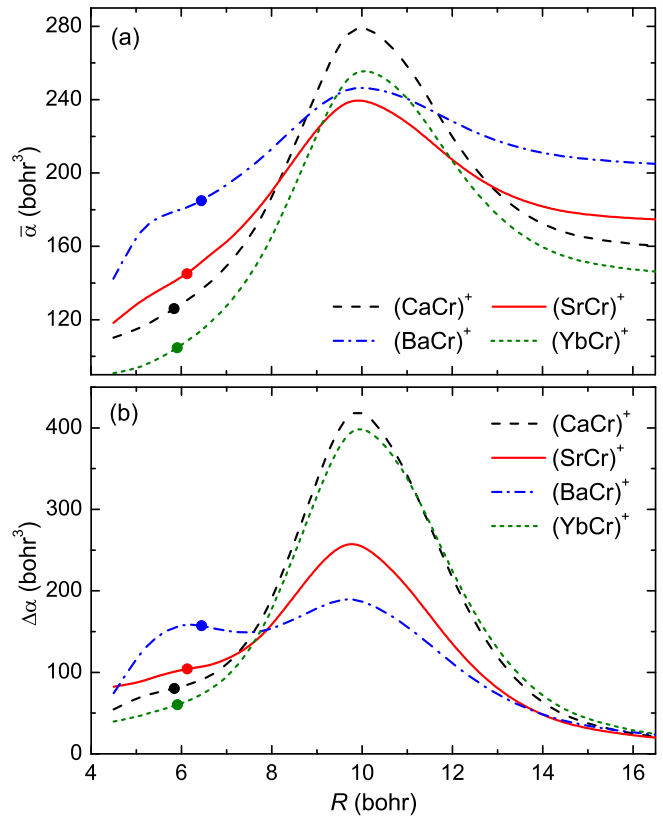}
\end{center}
\caption{(Color online) Average static electric dipole polarizability (a) and polarizability anisotropy (b) of the (CaCr)$^+$, (SrCr)$^+$, (BaCr)$^+$, and (YbCr)$^+$ molecular ions in the $X^6\Sigma^+$ electronic state as a function of the internuclear distance. The points indicate the values for the ground rovibrational level.}
\label{fig:pol_X}
\end{figure}

The  permanent electric dipole moments are calculated with respect to the center of mass, which is a natural choice for investigating the rovibrational dynamics. Their absolute values increase with increasing internuclear distance and asymptomatically approach the limiting cases where the charge is completely localized on the ion (the  dotted lines in Fig.~\ref{fig:dip}). 
The obtained dipole moments of the $a^8\Sigma^+$ and  $X^6\Sigma^+$ electronic states are similar.
The difference between the calculated values and the limiting cases is the interaction-induced variation of the permanent electric dipole moment or, in other words, the degree of charge delocalization.  This behavior is typical for heteronuclear molecular ions and implies that even  molecular ions in very weakly bound states have effectively a significant permanent electric dipole moment in contrast to neutral molecules. In fact, for heteronuclear molecular ions the value of the vibrationally averaged permanent electric dipole moment in the body-fixed frame with respect to the center of mass increases with decreasing binding energy as shown in Fig.~\ref{fig:vib_dip} for the investigated molecular ions in the $a^8\Sigma^+$ electronic state. For levels with binding energies below 1$\,$GHz the electric dipole moment exceeds 100$\,$D and reaches 1000$\,$D for the most weakly bound levels.
Thus, weakly bound molecular ions have a giant permanent electric dipole moment and present excellent prospects for controlling their dynamics with the nonresonant laser field.
 
The perpendicular $\alpha_{\perp}\equiv\alpha_{xx}=\alpha_{yy}$ and parallel $\alpha_{\parallel}\equiv\alpha_{zz}$  components of the static electric dipole polarizability tensor were obtained in the \textit{ab initio} calculations.
The average polarizability $\bar{\alpha}=(2\alpha_{\perp}+\alpha_{\parallel})/3$ and the polarizability anisotropy  $\Delta\alpha=\alpha_{\parallel}-\alpha_{\perp}$ of the (CaCr)$^+$, (SrCr)$^+$, (BaCr)$^+$, and (YbCr)$^+$ molecular ions in the  $a^8\Sigma^+$ and  $X^6\Sigma^+$ electronic states as functions of the internuclear distance $R$  are presented in Fig.~\ref{fig:pol_a} and Fig.~\ref{fig:pol_X}, respectively. 
The values for the ground rovibrational level are reported in Table~\ref{tab:spec}.

The polarizabilities show an overall smooth behavior
and tend smoothly to their asymptotic atomic values.
The interaction-induced variations of the polarizability are clearly
visible while changing the internuclear distance $R$ and are more pronounced for the $X^6\Sigma^+$ electronic state as compared to the $a^8\Sigma^+$ one.
The polarizability anisotropies for the rovibrational ground state
of the investigated molecular ions are slightly smaller than the ones of the alkali-metal-atom dimers~\cite{DeiglmayrJCP08} and chromium--closed-shell-atom molecules~\cite{TomzaPRA13a}. 
Here, we present the static polarizabilities which describe the interaction of the molecular ions with the far nonresonant laser field. When the shorter-wavelength laser field is applied the dynamic polarizabilities have to be used, which usually are larger but of the same order of magnitude as the static ones.

\subsection{Magnetic Feshbach resonances}

Fermionic alkaline-earth-metal and ytterbium ions, and bosonic chromium atoms do not have nuclear spin and therefore they do not possess any hyperfine structure, in contrast to 
bosonic alkaline-earth-metal and ytterbium ions, and fermionic chromium atom.  
Three types of magnetically tunable Feshbach resonances between the considered ions and atoms exist, depending on the structure of the ion and atom: (i)~Feshbach resonances between bosonic ions and fermionic atoms are of the same nature as those between two alkali-metal atoms~\cite{JulienneRMP10}. 
(ii)~Feshbach resonances between fermionic (bosonic) ions and fermionic (bosonic) atoms  
result from the interaction of the hyperfine structure of the atom (ion) with the electronic spin of the ion (atom), respectively~\cite{IdziaszekPRA09,IdziaszekNJP11,TomzaPRA15a}. (iii)~Feshbach resonances between fermionic ions and bosonic atoms are of the same nature as those between two bosonic chromium atoms, that is, they are due to the spin-spin interaction~\cite{WernerPRL05,PavlovicPRA05}. Here, we present example calculations for the Feshbach resonances of the type~(i). The resonances of the types~(ii) and~(iii) appear at smaller magnetic field strengths and their typical widths are smaller as compared to the type~(i)~\cite{TomzaPRA15a,WernerPRL05}.

\begin{figure}[t!]
\begin{center}
\includegraphics[width=\columnwidth]{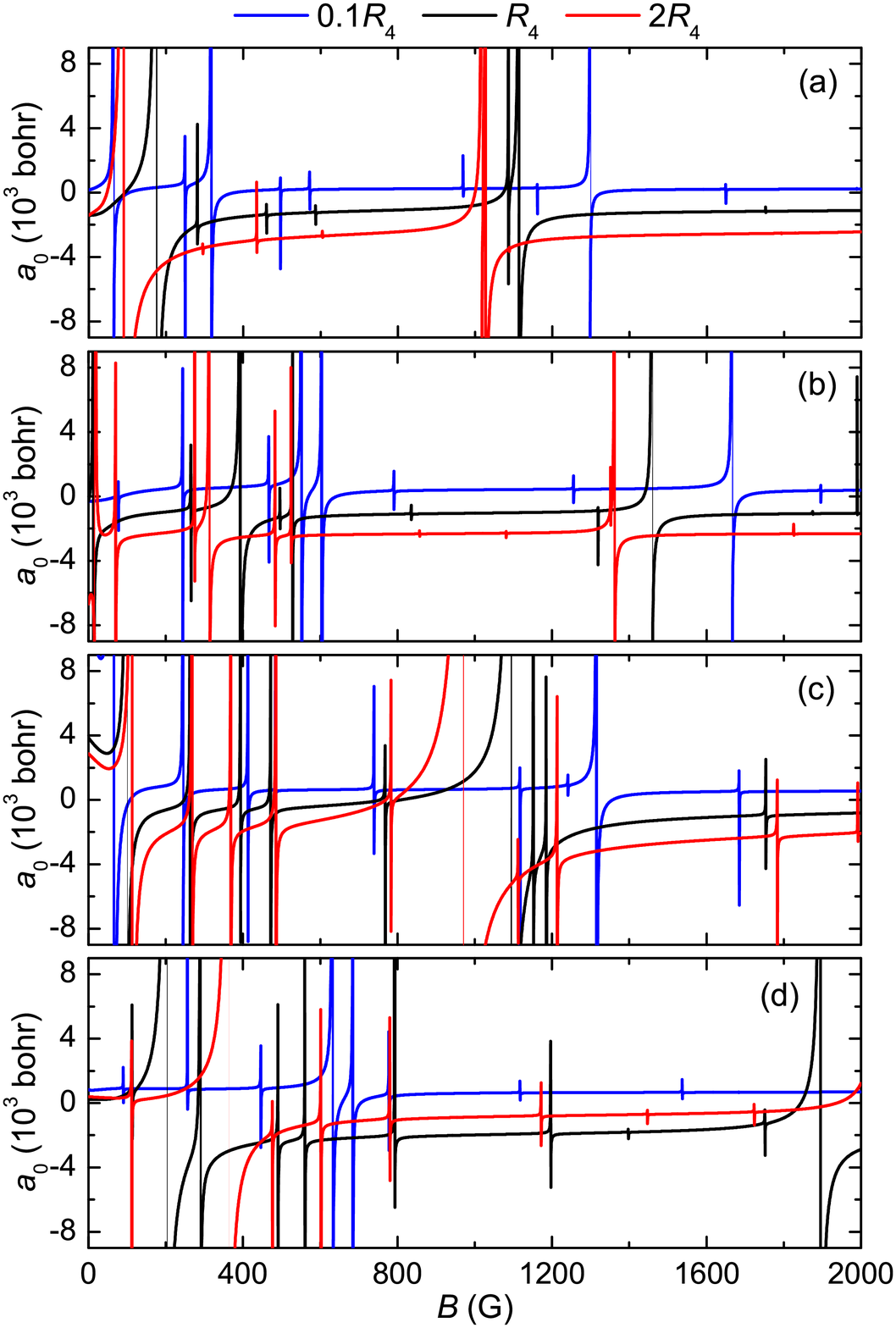}
\end{center}
\caption{(Color online) $s$-wave scattering length as a function of the magnetic field strength $B$ for collisions of the $^{53}$Cr~$(\frac{9}{2},-\frac{7}{2})$ atom with the $^{43}$Ca$^+$~$(4,-3)$~(a), Sr$^+$~$(5,-4)$~(b), Ba$^+$~$(1,1)$~(c), and Yb$^+$~$(3,-2)$~(d) ions at the collision energy of 100$\,$nK. Typical scattering lengths of $0.1R_4$, $R_4$, and $2R_4$ for the $X^6\Sigma^+$ electronic state and $-0.1R_4$, $-R_4$, and $-2R_4$ for the  $a^8\Sigma^+$ electronic state are assumed, respectively.}
\label{fig:FR}
\end{figure}

Positions and widths of Feshbach resonances depend on the background scattering length, the hyperfine structure, and the progression of weakly bound rovibrational levels just below the atomic threshold.   
Unfortunately, even the most accurate potential energy curves obtained in the most sophisticated \textit{ab intio} calculations do not allow one to predict accurately the scattering lengths for collisions between many-electron atoms. Thus, at present, it is impossible to determine all parameters of Feshbach resonances without \textit{a priori} experimental knowledge. Nevertheless, the general characteristics of Feshbach resonances  such as the density of resonances and typical widths can be learned by tuning the scattering lengths around the value of the characteristic length scale of the ion-atom interaction. This length scale is given by $R_4=\sqrt{2\mu C_4/\hbar}$~\cite{IdziaszekPRA09,IdziaszekNJP11} and amounts to 1878$\,$bohr for $^{43}$Ca$^++^{53}$Cr, 2212$\,$bohr for $^{87}$Sr$^{+}+^{53}$Cr, 2383$\,$bohr for $^{137}$Ba$^{+}+^{53}$Cr, and 2455$\,$bohr for $^{173}$Yb$^{+}+^{53}$Cr. The characteristic energy scale of the ion-atom interaction is given by $E_4=\hbar^2/(2\mu R^2_4)$ and amounts to 1.1$\,\mu$K for $^{43}$Ca$^++^{53}$Cr, 0.55$\,\mu$K for $^{87}$Sr$^{+}+^{53}$Cr, 0.41$\,\mu$K for $^{137}$Ba$^{+}+^{53}$Cr, and 0.37$\,\mu$K for $^{173}$Yb$^{+}+^{53}$Cr. For collision energies larger than $E_4$, higher partial waves contribute to the rates for elastic scattering and pronounced shape resonances, typical for ion-atom collisions, can be expected~\cite{TomzaPRA15a}.

Figure~\ref{fig:FR} shows the $s$-wave scattering lengths as functions of the magnetic field strength $B$ for ultracold collisions between the $^{53}$Cr~$(\frac{9}{2},-\frac{7}{2})$ atom and the $^{43}$Ca$^+$~$(4,-3)$, Sr$^+$~$(5,-4)$, Ba$^+$~$(1,1)$, and Yb$^+$~$(3,-2)$ ions, respectively. $(f,m_f)$ denotes the hyperfine state of the atom (ion), where $f$ is the total angular momentum of the atom (ion) and $m_f$ is its projection along a quantization axis defined by the external magnetic field $B$.
The results for three sets of the scattering lengths associated with the $X^6\Sigma^+$ and $a^8 \Sigma^+$ electronic states are presented. 
The same absolute values of $0.1R_4$, $R_4$, and $2R_4$ are assumed for both states with the positive sign for $X^6\Sigma^+$ and the negative sign for $a^8\Sigma^+$. 
The scattering lengths are set by scaling the potential energy curves with appropriate factors $\lambda$, $V(R)\to\lambda V(R)$,  taking values in the range of 0.99-1.01. 
Two families of resonances can be distinguished in Fig.~\ref{fig:FR}: broad ones and narrow ones. 
The broad resonances are caused by the hyperfine interaction in the ion, whereas the narrow resonances are due to the hyperfine interaction in the chromium atom.
The observed difference in the resonance widths results from the difference in the hyperfine coupling constant, which is much smaller for the chromium atom as compared to the ions.
The widths of the broad resonances are of the order of a few to several Gauss. The widths of the narrow resonances are relatively small, mostly below 1 G.
The density of the narrow resonances is higher than the density of the broad ones.
Resonances due to the dipolar spin-spin interaction, which couples $s$-wave channel to $d$-wave channel~\cite{PavlovicPRA10}, are unnoticeable in the scale of the plot.
To avoid potential losses due to the Zeeman relaxation in an atomic gas~\cite{PavlovicPRA05}, the chromium atoms in the lowest Zeeman and hyperfine state can be used.
When choosing the fully spin-stretched hyperfine state of the atom (ion), only broad (narrow) Feshbach resonances are present in the system.    

Finally, based on the above presented calculations for typical scattering lengths and example hyperfine states of the ions and atoms, one can expect the existence of at least one broad Feshbach resonance and many narrow Feshbach resonances at magnetic field strengths below 1000$\,$G in the case of the investigated type~(i) of ion-atom mixtures. More Feshbach resonances at smaller magnetic field strengths but with smaller widths are expected for the types~(ii) and~(iii) of systems.
The probability that the true potential energy curves correspond to very unfavorable values of the scattering lengths are rather small and such an unfortunate case can be remedied by changing the hyperfine state or isotope.

\section{Summary and conclusions}
\label{sec:summary}

We have proposed the Ca$^+$, Sr$^+$, Ba$^+$, and Yb$^+$ ions immersed in an ultracold gas of the Cr atoms as systems in which ion-atom interactions at ultralow temperatures can be controlled with magnetically tunable Feshbach resonances without charge transfer and radiative losses. 
We have investigated \textit{ab initio} properties of the $X^6\Sigma^+$ and $a^8\Sigma^+$ electronic states of the (CaCr)$^+$, (SrCr)$^+$, (BaCr)$^+$, and (YbCr)$^+$ molecular ions. Potential energy curves, permanent electric dipole moments, and static electric dipole polarizabilities have been calculated with the spin restricted open-shell coupled cluster method restricted to single, double, and noniterative triple excitations and the multireference configuration interaction method restricted to single and double excitations. Next, magnetically tunable Feshbach resonances have been analyzed. 

The ionization potential of the chromium atom is larger than the electron affinity of the alkaline-earth-metal and ytterbium ions. Therefore, the charge transfer process is energetically forbidden and thus the radiative losses observed in other experimentally investigated cold hybrid ion-atom systems can be avoided in the proposed mixtures. 
The other possible combinations are the Ba$^+$ ions and the Li atoms as well as 
the Sr$^+$ or Ba$^+$ ions and the Dy or Er atoms.
The avoidance of radiative charge transfer losses opens the way towards robust quantum simulations~\cite{BissbortPRL13} and computations~\cite{DoerkPRA10} with ultracold heteronuclear ion-atom systems.

Magnetically tunable Fesbhach resonances have been investigated for typical scattering lengths and example hyperfine states of the ions and atoms. The assumed ultracold collision energies require trapping ions in the dipole trap rather than in the Paul trap to avoid heating due to the rf-field-induced micromotion~\cite{HarterCP14}. The lack of the radiative charge transfer losses in the investigated systems should enable sympathetic cooling of ions to ultracold temperatures.
Presented calculations suggest the existence of useful Fesbhach resonances at moderate magnetic field strengths but the exact positions and widths of these resonances depend on the true values of the scattering lengths which have to be provided by experimental measurements.

The reduction of the two-body losses should also facilitate the formation, observation, and application of molecular ions. Magnetically tunable Feshbach resonances can be employed in the magnetoassociation~\cite{JulienneRMP10}, which can be followed by an optical stabilization to more deeply bound rovibrational levels. The photoassociation to excited electronic states should also be possible~\cite{JulienneRMP06a} but the actual pathways of this laser-induced association have to be determined. Additionally, the laser excitation of the ions or atoms enables radiative charge transfer thus allowing for the study of controlled chemical reactions in the investigated systems as well. 
In the absence of the two-body losses, it would be interesting to investigate the three-body ion-atom recombination process~\cite{HarterPRL12,HarteNatPhys13} 
\begin{equation}
X^+ +2\,\mathrm{Cr} \to (X\mathrm{Cr})^+(v,j) + \mathrm{Cr} 
\end{equation}
which leads to the formation of the molecular ions $(X\mathrm{Cr})^+$ in different rovibrational levels $(v,j)$. The control of this process with magnetically tunable Feshbach resonances both between ion and atoms as well as between atoms 
should be possible and give new insight into few-body physics. One can also think about investigating Efimov physics in the ion-atom-atom scenario~\cite{BraatenAP07}.

\begin{acknowledgments}
Financial support from the Marie Curie COFUND action through the ICFOnest program, the Foundation for Polish Science within the START program, the EU grants ERC AdG OSYRIS, FP7 SIQS and EQuaM, FETPROACT QUIC, the Spanish Ministry grant FOQUS (FIS2013-46768), and the Generalitat de Catalunya project 2014 SGR 874 is gratefully acknowledged.
\end{acknowledgments}

\bibliography{../0Bib/MT,../0Bib/ions}

\end{document}